\documentclass[apj]{emulateapj}
\setkeys{Gin}{width=0.48\textwidth}

\newcommand{\hmpc}{\ifmmode{h^{-1}\,\hbox{Mpc}}\else{$h^{-1}$\thinspace Mpc}\fi}

\newcommand{\kms}{\ifmmode{\,\hbox{km\,s}^{-1}}\else {\rm\,km\,s$^{-1}$}\fi}
\newcommand{\msun}{{\rm\,M_\odot}}

\begin{document}
\title{The Pal~5 Star Stream Gaps} 
\shorttitle{Pal~5 Gaps}
\shortauthors{Carlberg, Grillmair \& Hetherington}
\author{R. G. Carlberg\altaffilmark{1} }
\author{C. J. Grillmair\altaffilmark{2} }
\author{Nathan Hetherington\altaffilmark{1} }
\altaffiltext{1}{Department of Astronomy and Astrophysics, University of Toronto, Toronto, ON M5S 3H4, Canada {carlberg,hetherington}@astro.utoronto.ca }
\altaffiltext{2}{Spitzer Science Center, 1200 E. California Blvd., Pasadena, CA 91125, USA, 
carl@ipac.caltech.edu}


\begin{abstract}
Pal~5 is a low mass, low velocity dispersion, globular cluster with spectacular tidal tails. We use the SDSS DR8 data to extend the density measurements of the trailing star stream to 23 degrees distance from the cluster, at which point the stream runs off the edge of the available sky coverage. 
The size and the number of gaps in the stream are measured using a filter which approximates the structure of the gaps found in stream simulations.  
We find 5 gaps that are at least 99\% confidence detections  with about a dozen gaps at 90\% confidence.  The statistical significance of a gap is estimated using bootstrap re-sampling of the control regions on either side of the stream. 
The density minimum closest to the cluster is likely the result of the epicyclic orbits of the tidal outflow and has been discounted.  To create the number of 99\% confidence gaps per unit length at the mean age of the stream requires a halo population of nearly a thousand dark matter sub-halos with peak circular velocities above 1 \kms\ within 30kpc of the galactic center. These numbers are a factor of about 3 below cold stream simulation at this sub-halo mass or velocity, but given the uncertainties in both measurement and more realistic warm stream modeling, are in substantial agreement with the LCDM prediction.
\end{abstract}

\section{INTRODUCTION}
\nobreak

\citet{Abell:55} cataloged the globular cluster Pal~5 in the National Geographic Society - Palomar Observatory Sky Survey, noting that Baade and \citet{Wilson:55} had each independently discovered the object, which Wilson listed as the Serpens cluster.
\citet{Arp:65} assigned Pal~5 the lowest possible \citet{SS:27} concentration class. The cluster also has an unusually low luminosity \citep{SH:75}.  The quantitative King model \citep{vonHoerner,King:62,King:66} concentration parameter is a notably, but not uniquely, low 0.52 \citep{KHHW:68,Woltjer:75,Harris}.  

In the early photographic data the density profiles became uncertain toward the tidal radius as the cluster sinks into the distribution of foreground and background stars in the galaxy. \citet{GFIQ:95} acquired new, multi-degree, photographic data and introduced photometric selection procedures to screen out background stars, yielding the important discovery that near the tidal radius the stars were distributed in asymmetric clouds with a ``striking resemblance" to the expected tidal tails. 
The depth and photometric uniformity of the Sloan Digital Sky Survey \citep{SDSS}  have provided increasingly good views of the tidal tails of Pal~5.  In the SDSS commissioning data the Pal~5 tails were detected out to a distance of about $\pm$1.3\degr\ from the cluster \citep{Odenkirchen:01}. 
Increased sky coverage of the SDSS and the introduction of a statistically optimal matched filter technique \citep{Rockosi:02} allowed the trailing tail to be traced to about 6.5\degr\ from the cluster \citep{Odenkirchen:03}.  
Their resulting detection of density variations along the stream was a major new development. SDSS DR4 data improved both the sky coverage and the photometry, allowing \citet{GD:06} to trace the trailing tail to about 16\degr\ from the cluster and the leading tail to about 6\degr, where it runs off the edge of the current survey area.

\begin{figure}
\begin{center}
\includegraphics[scale=0.5]{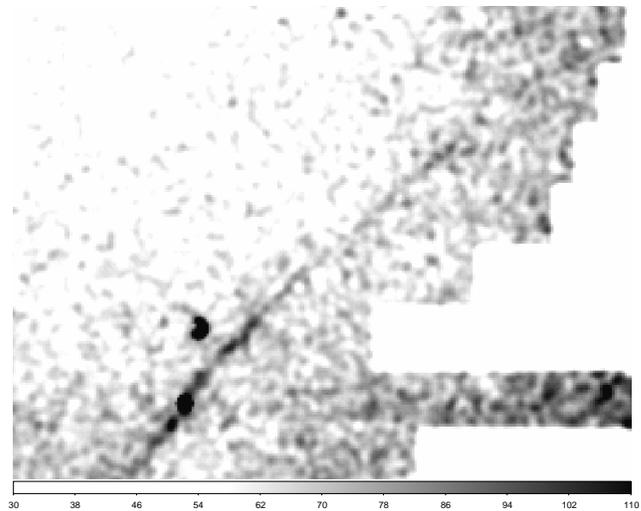}
\end{center}
\caption{The match filtered star densities in the region of the Pal~5 stream in the SDSS $\lambda$ and $\eta$ co-ordinate system.  The raw image has been smoothed with a 3 pixel Gaussian. The object above the stream is the foreground cluster M5.}
\label{fig_stream}
\end{figure}

\begin{figure*}
\begin{center}
\includegraphics[angle=0,bb= 44 318 572 480,scale=2.1]{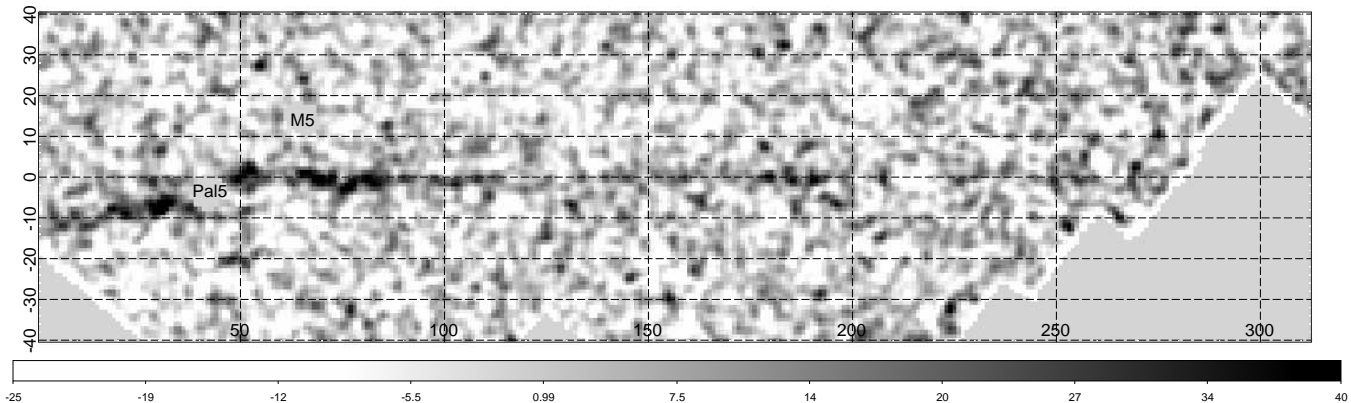}
\end{center}
\caption{The matched filtered star map of the Pal~5 field, with Pal~5 and the foreground M5 cluster masked out. To remove the varying background, the masked image has been smoothed over 4\degr\, subtracted from the original image, and then smoothed with a 2 pixel, or, 0.2\degr\ Gaussian. The analysis is conducted on the original uncorrelated pixels. We have made no attempt to straighten the southern part of the stream, left of the cluster in this image.}
\label{fig_image}
\end{figure*}

Current data shows Pal~5 to be one of the lower luminosity\citep{Harris} and lowest velocity dispersion, $1.1 \pm 0.2 \kms$ \citep{Odenkirchen:02}, globular clusters known. 
The long, thin, cool tidal tails \citep{GD:06,Odenkirchen:09} have significant density variations along their length. Consequently it has become a key system for the study of the small scale dynamics of the galactic halo with an emphasis on the physics behind the origin of the star stream density variations.

In a series of papers \citet{Kupper:08,Kupper:10, Kupper:12} use both dynamical analysis and precise n-body simulations to demonstrate that stars leaving a globular cluster at low velocity through the Lagrange points execute coherent epicyclic orbits which lead to pileups at the low points of the cycloids and low density regions between.  Their detailed n-body simulations show that the periodic structures are present for general cluster orbits, but that the range in phase angles emerging through the Lagrange points cause the structures to blur out with distance down the stream and that all quantities vary as the cluster and stream orbit around the galaxy. 

Dwarf galaxies also produce star streams although they do not have the two and three body interactions that provide a nearly continuous flow of stars to the zero energy surface in a star cluster.  The possibility of dynamical instabilities \citep{CQ:11} that could create gaps if they become nonlinear has been raised. However,  \citet{SM:11} demonstrate that dynamical instabilities in the lengthening star streams in strong tidal fields are not likely to be present. For globular clusters, stream dynamical instabilities are not seen in the n-body simulations \citep{Kupper:12}.

Cosmological n-body simulations from LCDM initial conditions find that galactic halos will have approaching 10\% of their mass in vast numbers of orbiting sub-halos \citep{VL1,Aquarius,Stadel:09} which will act on star streams to locally deflect and heat them \citep{Ibata:02,JSH:02,SGV:08,Carlberg:09} and induce visible gaps \citep{YJH:11,Carlberg:12}.  The rate at which gaps are randomly created is approximately constant in time in a relatively slowly evolving, low redshift, galactic halo. Whereas the epicyclic pileups decrease with distance down the stream, sub-halo induced gaps increase in number and become better defined with age or distance down the stream.

Here we take advantage of the SDSS DR8 data \citep{DR8} in the Pal~5 region to better extract the density along the trailing stream.  We use the adjoining sky regions to estimate the background and to estimate the errors in the density measurement.  To characterize the gaps we develop a gap-finding filter. The resulting measurements of gap sizes and numbers are discussed in relation to the ideas for epicyclic lumps and dark matter sub-halo induced gaps as a test of the LCDM dark matter sub-halo predictions.

\section{Pal 5 Data Analysis}

The SDSS Data Release 8  \citep{DR8} has improved photometric uniformity over earlier releases along with a modest increase in sky coverage in the Pal~5 region. The stars are counted in 0.1\degr\ square pixels using the well-developed matched filter technique \citep{Rockosi:02,Grillmair:09,Grillmair:11}. The pixel values are star counts filtered on the basis of their agreement with old, low metallicity stellar isochrones, weighted with the $\omega$ Cen luminosity function \citep{deMarchi:99}  using $g, r$ and $i$  photometry  corrected for reddening. The binned, weighted star count data in the SDSS co-ordinate system are shown in Figure~\ref{fig_stream}. 

\subsection{The Extracted Stream}

The northern, trailing, part of the stream is slightly curved along its length \citep{GD:06}, as shown in Figure~\ref{fig_stream}.  The stream centerline is defined first by rotating the stream to a horizontal x axis then removing the residual variation with a spline function along the stream.
We define the centerline of the stream with an eye fit, then iterate the positions but find that the improvement in the total integrated luminosity of the stream is not significant for a reasonable choice of the stream centerline. 
For each pixel we calculate xy co-ordinates relative to the centerline with the $y$ (vertical) component set to zero.
The pixels are maintained as statistically independent measurement points for the analysis. However, for the purpose of illustration, the data are interpolated back onto a square 0.1\degr\ grid, a 4\degr\ Gaussian smoothed image subtracted, and the difference is then smoothed with a 0.2\degr\ circular Gaussian. The two clusters are masked out.  The image shown in Figure~\ref{fig_image} places the fitted centerline at y=0 pixels, which leads to the cluster being positioned at [x,y]=[42,-5]. At the distance of Pal~5, 23.2 kpc \citep{Harris}, each 0.1\degr\ pixel subtends approximately 0.0405 kpc.  

The stream is clearly detected over the entire length of the image, 42 to 274 pixels, an angular distance of 23.2\degr\ from the cluster. The projected  total length of the northern star stream is 9.4 kpc. The background rises slightly towards the end of the stream. We note that there is a negative density dip around x=215 pixels, with no clear source in the star counts or dust map.   

\begin{figure}
\begin{center}
\plotone{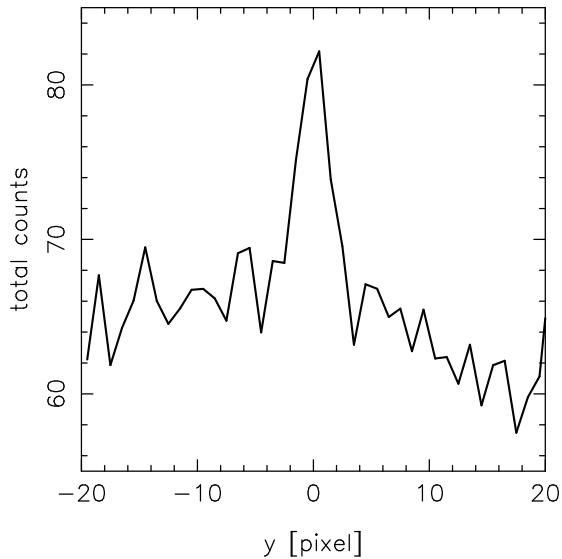}
\end{center}
\caption{The density profile transverse to the stream averaged over x=[40,272]. }
\label{fig_width}
\end{figure}

The bilinear interpolation and filtering used to create Figure~\ref{fig_image} introduces correlations between the pixels, so all analysis is done using using the non-integer xy co-ordinates of the original pixels relative to the centerline.  Figure~\ref{fig_width} shows the density profile summed along the stream in 1 pixel bins transverse to the centerline.  The FWHM of the stream is approximately 3 pixels, equal to the 0.12 kpc that \citet{Odenkirchen:03} estimate, but here over a nearly four times longer extent. We will normally do the analysis over a strip of total width of $\pm2$ pixels from the centerline, a full width of 4 pixels, capturing more than 95\% of the stream density, which we use as the stream density profile in all of the analysis. The background is estimated from points on either side of the stream no closer than 5 pixels to the centerline, generally using two strips of five times the stream width, or 10 pixels. Using adjacent regions of 3 to 7 times the stream width changes the results about 10\% or less.

\begin{figure}
\begin{center}
\plotone{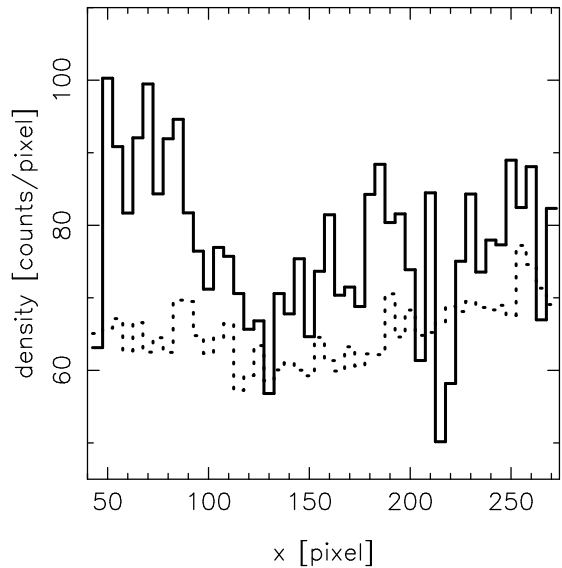}
\end{center}
\caption{The density along the stream (solid line) and in the surrounding background region (dotted line) in 0.5\degr, or 5 pixel, bins. The background region is 4 times the width of the stream to reduce noise. The cluster is at x=42. }
\label{fig_p1}
\end{figure}

The two-dimensional map of Figure~\ref{fig_image} shows that the stream has small deviations from the centerline along its length, as was previously noted in \citet{GD:06}, so some of the width is due to centerline wandering, not true width. We see no clear evidence for a systematic variation of stream width along its length. The full 2D map is potentially an extremely powerful tool in the analysis of the physical cause of variations in the stream density as the image signal to noise improves with better data.

\begin{figure}
\begin{center}
\plotone{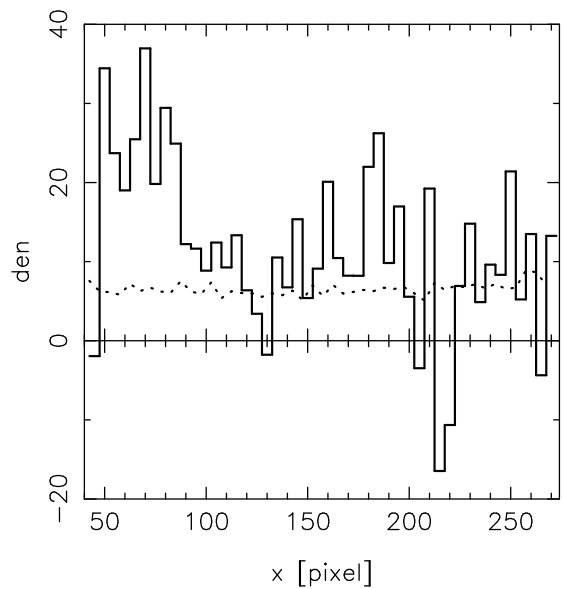}
\end{center}
\caption{The background subtracted density along the stream (solid line) and the error estimate (dashed line) in 0.5\degr, or 5 pixel, bins.}
\label{fig_dene}
\end{figure}

\subsection{The Stream Density Profile and Its Errors}

The density along the stream and in the two adjacent regions used to estimate the background,  binned over 5 pixels along the stream to suppress noise, is shown in Figure~\ref{fig_p1}.  There is a slight rise in the background near the visible end of the stream, but it shows no small scale systematic deviations from the stream which would induce systematic errors that could be confused with gaps. 
The stream itself is detected at high significance along its entire length. The subtracted density profile along with the error estimate, both in the 5 pixels bins, is shown in Figure~\ref{fig_dene}. 

\subsubsection{Random Errors}

Constructing a stream density error estimate is crucial to the statistical tests for the significance of gaps. Our fundamental assumption is that the background level at the location of the stream and the stream error properties can be derived from the averages and standard deviations of the pixels in the regions to either side of the stream. The background itself is simply the average of the two sides. 

The standard deviation  in a background pixel parallel to stream location $i$ is calculated from pixels spread over a total width that is $N$ times the width of the stream. Statistically, the background will have a standard deviation of $\sigma_i/\sqrt{N}$, where $\sigma_i$ is the standard deviation in a region the width of the density stream.
Accordingly,  we take the error in the measurement of the derived stream density to be $\sqrt{1+N}$ times the value measured in the background region, where we have made use of the fact that the stream is a relatively small, $\simeq20\%$, over-density above the background, which in quadrature would change the errors about 10\%.
The exact value of $\sigma_i$ varies from pixel to pixel due to noise, but in practice it is convenient that there is no systematic trend along the stream, as shown in  Figure~\ref{fig_dene}. 

In detail, the stream density error estimate is constructed using the background region a minimum of 5 pixels from the centerline of the stream and generally five times the width of the stream on either side of it, all normalized to give the value appropriate for the stream itself.
 We have performed the analysis with varying background regions from 3 to 7 times the width of the stream and see no differences above those expected from the estimated noise level. 
The individual pixel density values in the background region, with a quadratic polynomial subtracted to remove the local mean, provides a basis for constructing gap-free bootstrap realizations of the stream density. 
That is, we construct random gap-less streams with a noise level the same as the real stream, using the background density profile along the stream, with the mean subtracted, scaled by $\sqrt{1+N}$ and then create bootstrap samples using random drawing with replacement. 

\begin{figure}
\begin{center}
\includegraphics[angle=-90,scale=0.7]{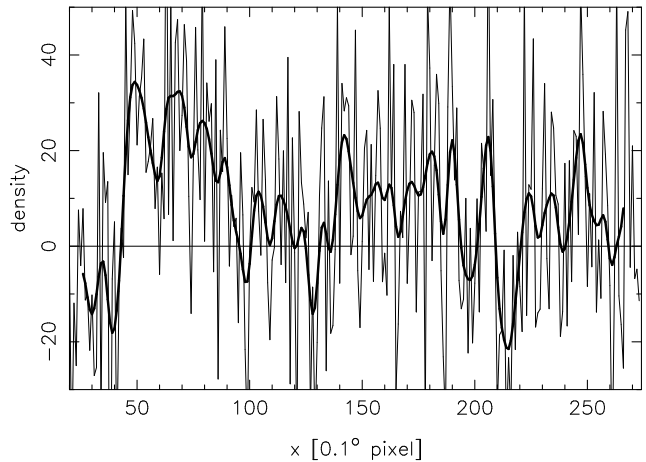}
\end{center}
\caption{The background subtracted density along the stream in 0.1\degr\ pixels (thin line) and filtered with a 0.2\degr\ width Gaussian (thick line). The Pal~5 cluster is at x=42. }
\label{fig_density}
\end{figure}

The background subtracted density in the 0.1\degr\ bins that we use for analysis is shown in Figure~\ref{fig_density}, along with a 0.2\degr\ Gaussian filtered version of the density to give better signal-to-noise for the purpose of the figure. The density profile in the first 6\degr, from x=40 to 100 recovers the features first reported in \citet{Odenkirchen:03}. 

\begin{figure}
\begin{center}
\includegraphics[angle=-90,scale=0.7]{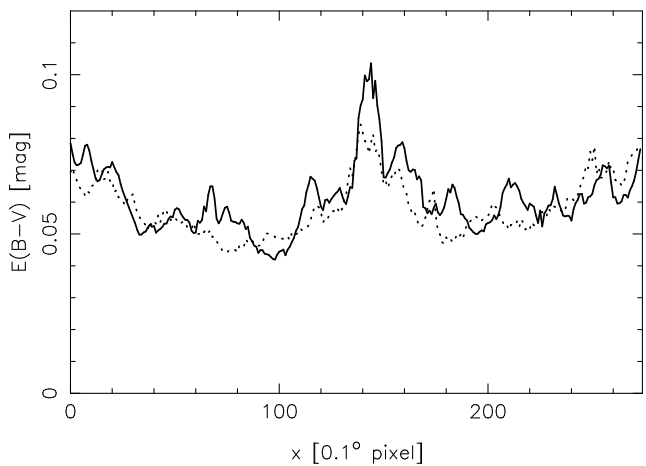}
\end{center}
\caption{The reddening along the stream (solid line) and in the background region (dashed line) in 0.1\degr\ pixels.}
\label{fig_dust}
\end{figure}

\subsubsection{Extinction Systematic Errors}

Figure~\ref{fig_dust} shows the  E(B-V) reddening values of \citet{SFD:98} measured along the location of the centerline of the stream and the adjacent background regions. The reddening along the stream is typically slowly varying with a mean near 0.05 mag, requiring brightness corrections of 0.15 mag or less. The extinction in the region we use for background is generally within 0.01 mag of the value along the stream. 

If the star counts with magnitude, $m$, were as steep as the Euclidean $d\log{(N)}/dm \propto 0.6 m$, an uncorrected uniform distribution would have extinction induced density variations of $\sim$23\% at most, and more typically about 7\%. Because the extinction correction has been made and using the background to stream extinction differences of 0.01 mag or less as an estimate of the error the residual variations in number will be less than one percent, about a factor of ten less than the statistical errors. The reddening and extinction corrections introduce weak correlations between the pixels, since the originating data  \citep{SFD:98} are provided in 0.32\degr\ pixels, more than three times larger than the star count pixels. We conclude that any systematic error due to dust variations is less than 1\%, hence is more than a factor of ten below the random errors.

\subsubsection{Random Error Consistency Test}

A simple $\chi^2$ test shows that the Pal~5 stream has density variations far above the noise.
The $\chi^2$ per degree of freedom for the data shown in Figure~\ref{fig_density} gives $\chi^2/\nu=2.13$ for 224 degrees of freedom, which indicates very high significance variations, that is, less than a part in $10^6$ that the density variations are due to noise . 

Along the stream we might expect the density of nearby points to be correlated. The auto-correlation of the mean subtracted density profile, $\delta d$, as a function of lag, $\sigma^2(\Delta x) = \int \delta(x)\delta(x+\Delta x) \,dx$, gives a cross-correlation of adjacent 0.1\degr\ pixels of  0.2, which is not significant. Furthermore, the correlation analysis, even when restricted to the first 6\degr, reveals only statistically weak evidence for periodic structure, which would be expected if the tidal cycloids dominated the density structure along the stream.

The auto-correlation function away from the initial peak provides an independent check on the size of the errors in the density distribution. The values of the correlation function away from the peak near zero lag are products of independent random variables whose population distribution has a standard deviation that is equal to the total variance of the data times the square root of the number of elements in the data. For the data range [50,274] the auto-correlation distribution gives the variance per element to be 28.6. The measured  $\chi^2/\nu=2.13$ and the point-by-point estimate of the error over this range is 14.16. Therefore $\chi^2/\nu$ estimates the variance per element to be 30.1. The two independently determined values agree within 5\%, which requires that the independently estimated $\sigma_i$ values used in the $\chi^2$ calculation be correct.
The agreement between two different methods to calculate the error and total variance gives us confidence in our overall error analysis. The next problem is to assign some fraction of the total variance to a systematic signal of stream gaps.

\section{Gaps in the Stream}

\begin{figure}
\begin{center}
\includegraphics[angle=-90,scale=0.7]{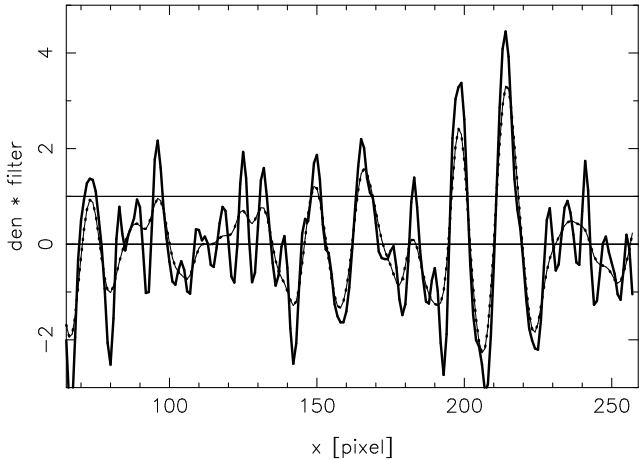}
\end{center}
\caption{The density profile as filtered with the $w_1$ (dotted thin line) and $w_2$ (solid line) filters for gaps of width 5 pixels. Both curves have the same normalization, the standard deviation of the $w_1$ filter.}
\label{fig_gap5}
\end{figure}

Sub-halo induced gaps will occur at random locations along the stream with the most distant, oldest, part of the stream expected to have relatively more gaps. Most gaps should be close in size to the smallest gaps that can survive the various blurring motions of the orbit, since the low mass sub-halos Narrow gaps should be more numerous because of the steep rise of sub-halo numbers with decreasing mass. Since mass is conserved, gaps in the streams must have approximately compensating positive and negative densities relative to a larger scale mean, although this assumption depends on measuring the density to sufficiently large distance from the stream centerline to include all scattered stars. The assumption will work best for smaller gaps.  

\subsection{Gap Filters}

A procedure to find gaps is to convolve a function that approximates the expected shape with the density data and identify peaks in the convolution. We use the functions $w_1(x) = (x^6 -1)\exp{(-1.2321 x^2)}$ and
$w_2(x) = (x^8 -1)\exp{(-0.559 x^4)}$, 
which rise at x=0 from -1 through zero near x=1 and then asymptotically back to zero, with integrals of (near) zero over [-3,3]. These filters have horns on either side of a trough, approximating the shape of the density gaps found in the simulations of \citet{Carlberg:12} with $w_2(x)$ having a much flatter floor and sharper horns. The whole procedure is a form of wavelet analysis. As illustrated in Figure~\ref{fig_gap5} the $w_2$ filter generally separates close peaks better, so we preferentially quote the results it gives.

We filter the density field with the two gap filters scaling them from 1 pixel up to 50 pixels in size and then find the peaks of the filtered field, corresponding to gaps in the density, and tabulate their heights and locations. To assign a confidence level to a given peak we run the same filters on our gap-less bootstrap samples constructed from the background regions. The gap filters have the important property that they are symmetric with zero mean, so the mean value nor a linear slope of the data within a filter makes no difference to the outcome. 
Although not very important for these data, we use a low order polynomial fit to the background density to subtract the slowly varying mean background which would otherwise excessive variance in the bootstrap samples.
We usually generate 100,001 bootstrap samples using the perl rand function, with randbit reported as 48. The heights of the peaks in the bootstrap samples are sorted and the heights which encompass, say 99\%,of the sample, defines the 99\% confidence levels as a function of the width of our gap filters. 

\subsection{Gap Statistics}

\begin{figure}
\begin{center}
\includegraphics[angle=-90,scale=0.7]{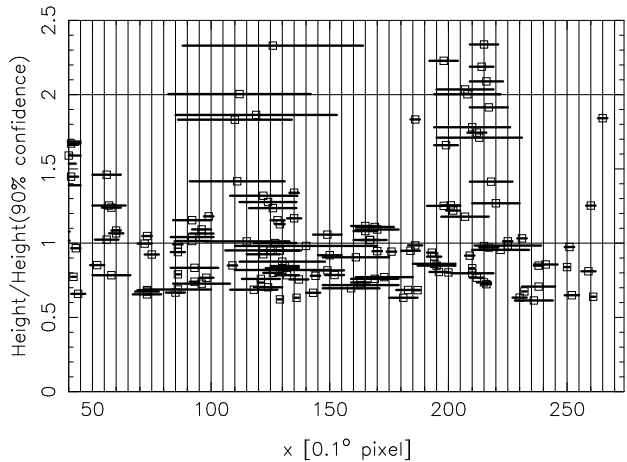} 
\end{center}
\caption{The gap-finder peaks along the stream, plotted relative to the 90\% confidence level for the $w_2$ filter.}
\label{fig_manygaps}
\end{figure}

\begin{figure}
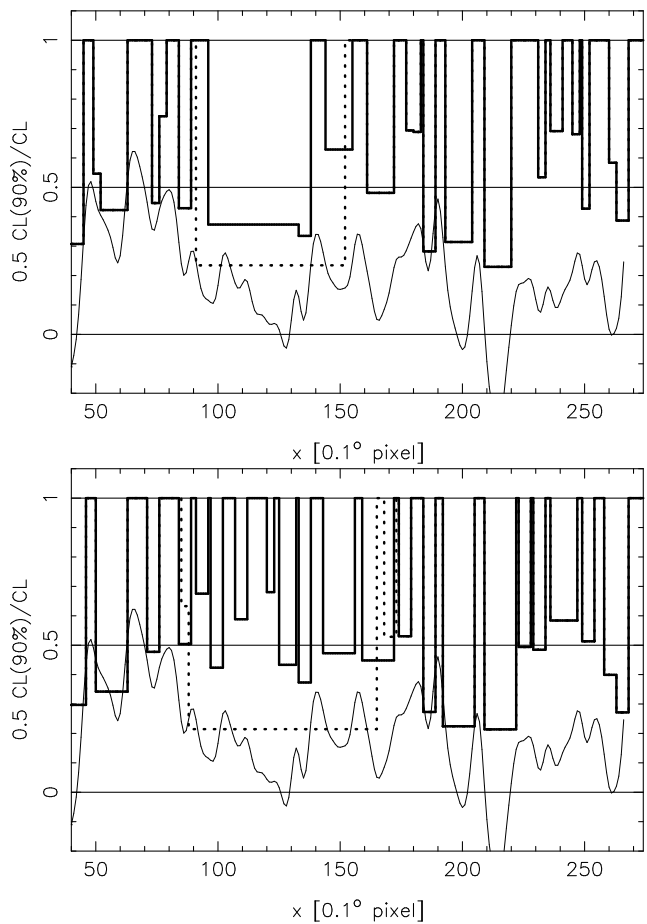

\begin{center}
\includegraphics[angle=-90,scale=0.7]{Figure_10a.eps} 
\includegraphics[angle=-90,scale=0.7]{Figure_10b.eps}
\end{center}
\caption{The gaps reconstructed from the highest significance gap filter peaks for gaps 1.8\degr\ and smaller (solid) and all gap sizes (dashed). The top plot is for the $w_1$ filter and the bottom for the $w_2$ filter. See text for details.
The plotted depth of the most significant gap at any location  is 0.5 CL(90\%)/CL, where CL is the confidence level.  Only the 4 and 5 highest confidence gaps for filters 1 and 2, respectively, survive at 99\% confidence The light line is the measured density of Fig.~\ref{fig_density}. }
\label{fig_recon}
\end{figure}

The whole range of gaps and their confidence levels is shown in Figure~\ref{fig_manygaps} for the $w_2$ filter scaled from 0.1\degr\ to 5.0\degr.  The plot shows all gaps above 67\% confidence. We note that most peaks are fairly stable, but that the peaks in the x=80 to 140 region merge together a set of narrow peaks into one very wide peak.  The origin of this behavior is visible in Figure~\ref{fig_density} where the density in this region becomes sufficiently low that the ``one big gap" interpretation is plausible. However the confidence level assigned to this large gap is sensitive to the details of the filter shape as shown in the two panels of Figure~\ref{fig_recon}. 


To illustrate the outcome of the filtering we schematically reconstitute the stream from the gap analysis. We sort the gaps by confidence level and insert gaps in a $d(x)=1$ distribution with an illustrative depth of $0.5 CL(90\%)/CL$ at the location of the highest significance gap at any location. A lower significance gap within a higher significance gap is ignored. The outcome using gaps of 1.8\degr\ (750~pc) and smaller is shown in Figure~\ref{fig_recon}. The wider filters merges a set of lower confidence peaks around x=120 into one. Although the graph is somewhat schematic the reconstituted gaps can be associated with features in the original data and with depths proportional to their statistical confidence.

To count gaps we use the highest significance gap at any location ignoring any lower significance gap(s) within its range.
  In the case illustrated we count 14(10) gaps with 90\% confidence or greater for the $w_1$($w_2$) filter.   The region of negative density around x=215 may be a problem in the star count analysis so, if present, we remove it from the peak count to reduce the counts by one. The fairly strong gap at x=55 is likely a tidal feature, discussed below, that we also remove if present. Therefore, after discounting, we find at 90\% confidence 12 (8) peaks. 
At 99\% confidence we find 4(5) peaks. Moving to a very high confidence level, we find 1(2) peaks (after discounting) at 99.99\% confidence. We will prefer the $w_2$ filter results as better matched to the shape of gaps in our simulations and more stable counts. Our rates will conservatively use the 99\% confidence counts, and use a multiplicative factor of two error, that is a range from 10 to 2, with 5 gaps as the preferred value.

\subsection{Mean Stream Density}

\begin{figure}
\begin{center}
\includegraphics[scale=0.9]{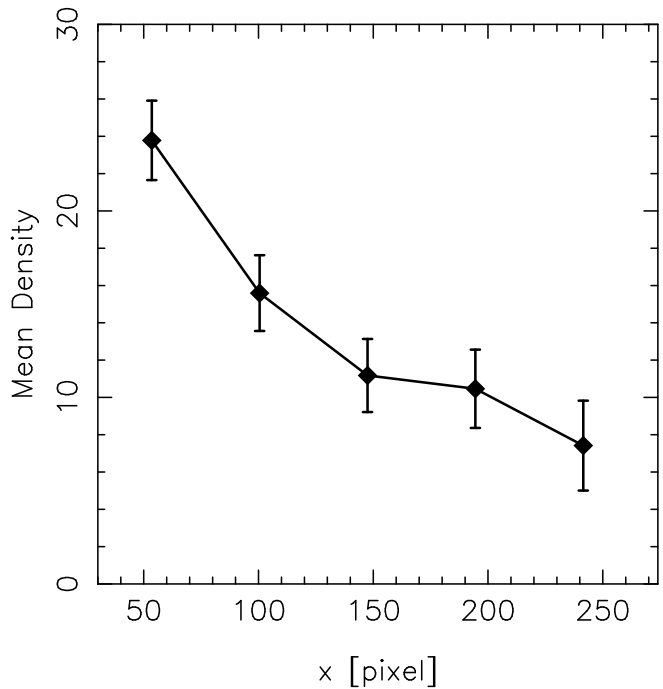}
\end{center}
\caption{The stream density in bins of 4.7\degr.}
\label{fig_meandensity}
\end{figure}

The mean density of the stream over 4.7\degr\ is shown in Figure~\ref{fig_meandensity} which illustrates the declining density with distance from the progenitor.   The decline is not likely to be a result of an increase of the tidal field, since at  roughly 7 Gyr old the Pal~5 stream has made several complete orbits. The mean density decline could be due to a changing mass loss rate from the cluster, but, simulations generally expect a fairly constant of even declining mass loss rate with time \citep{Kupper:10,DOGR:04}. A relation to gaps is that since the most distant part of the stream is the oldest that material that was in the gaps moves sufficiently far from the stream that it is not captured in our density measurement which assumes a constant width along the stream. 
Of course it is quite possible that the observed stream exhibits several of the effects combined.

\begin{deluxetable}{rr rrr rrr}
\tablecolumns{8}
\tablewidth{0pc}
\tablecaption{Observed Stream Gap Statistics}
\tablehead{
\colhead{Stream} & \colhead{Gaps}   & \colhead{Length} & \colhead{Width}   & \colhead{Age/2} & \colhead{$R_{GC}$}& \colhead{$n/n_0$} & \colhead{$\mathcal R_\cup$} \\
 & \# & kpc & kpc & Gyr  &kpc & &  kpc$^{-1}$  \\
 & &  &  &   & & & Gyr$^{-1}$  \\
}
 \startdata 
M31 		& 12 	& 200 	& 5 			& 5			& 100	&6	 	& 0.012\\
Pal~5		& 6	& 8.1		& 0.12		& 3.5		& 19		&22		 & 0.17\\
EBS			& 8		& 4.7		& 0.17		& 3.5		& 15 		&24 	&0.49	\\
Orphan 		& 2 		& 30	 	&1.0 			& 1.8		& 30		&17 	&0.037\\
\enddata
\end{deluxetable}

\begin{figure}
\begin{center}
\includegraphics[angle=0,bb= 41 29 505 515]{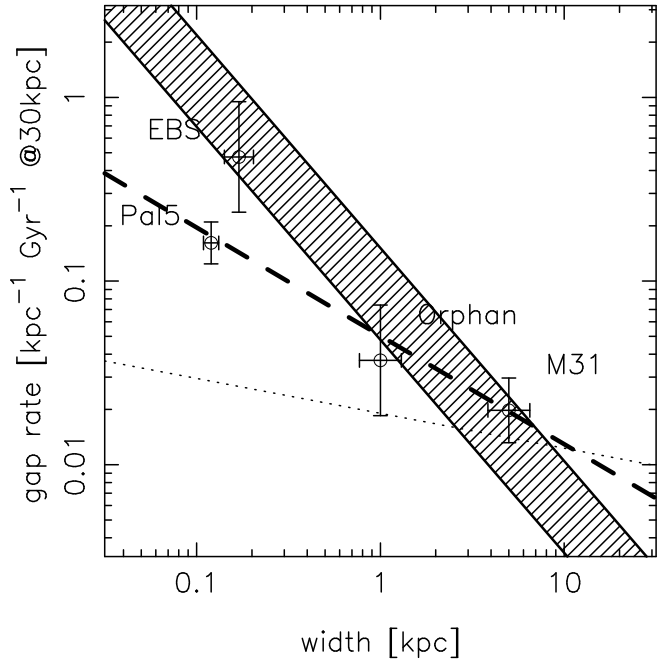} 
\end{center}
\caption{The gap rate-width relationship with all quantities scaled to 30 kpc radial distance from the galactic center. The dashed line is a variance weighted fit to the data. The dotted line is a model in which the number of halos varies as $M^{-1.8}$ normalized to have the same number of $M>10^9\msun$ sub-halos found for the $M^{-1.9}$ fit of simulations. The hashed region is an estimate of the theoretical uncertainties for the cold-stream calculation.}
\label{fig_gapwidth}
\end{figure}

\section{Gap Formation Processes} 

\subsection{Tidal Density Variations}

\citet{Kupper:08} find that the distance from the cluster to the first and subsequent epicyclic pileups is $x_L 4\pi \Omega(4\Omega^2 - \kappa^2)/\kappa^3$, where $x_L$ is the distance from the cluster center to the Lagrange point and $\Omega$ and $\kappa$ are respectively the rotational and epicyclic frequencies.  The Lagrange point is calculated as $x_L^3=GM /(4\Omega^2-\kappa^2)$, where $M$ is the cluster mass. 
For a locally flat rotation curve $\kappa=\sqrt{2}\Omega$ in which case the distance between epicyclic pileups is $2\sqrt{2}\,\pi x_L$. \citet{DOGR:04} estimate the tidal radius with $r_t^3 = GMR^2/v_c^2$ where $R$ is the galactocentric radius and $v_c$ is the galactic circular velocity at the cluster location, deriving $r_t$  to be 0.054 kpc. The Lagrange point distance for a flat rotation curve is $x_L^3= r_t^3/2$. Consequently $x_L=$ 0.043 kpc for Pal~5.  The epicyclic pileups at the base of the cycloidal type motion are then spaced 0.38 kpc apart which is 9.4  of our 0.1\degr\ pixels.  These calculations apply to a circular orbit, whereas Pal~5's orbit is fairly eccentric, for instance as modeled in \citet{Kupper:12} or \citet{MB:12}, but the same general considerations apply. In an eccentric orbit the spacing of the tidal stream density peaks and valleys vary as the stream moves around in the orbit. 

There are two types of tidal features worth seeking. The easiest are the  lumps expected at the cluster position, 42 pixels, plus the cycloidal distance, 9.4 pixels, or 51 pixels. There certainly is a density excess which peaks around 53 pixels, see Figures~\ref{fig_image} and \ref{fig_density}. There is a deficiency, not excess, at 60 pixels, but a clear excess at 70 pixels. There are also deviations from the centerline in (our) y direction, the most prominent deviation being at about 32 pixels from Pal~5, or 1.3 kpc. There may be a similar deviation in the same direction at half the distance, 16 pixels. 
These features seem sufficiently regular and of approximately the right character to be understood as the expected epicyclic pileups, but the distances are about a factor of two larger than the simple circular orbit theory predicts, which could reflect the complex orbital and tidal history of this unusual low concentration cluster \citep{Kupper:12}. We discount the gap at x=58 pixels as more likely to be a cycloidal orbit gap than a sub-halo induced gap.

\subsection{Gaps and Sub-Halos}

Assigning the gap at x=58 to the epicyclic motion we remove that gap from our counts. Using the 99\% confidence count of gaps we find 5 gaps. We estimate the age of the oldest part of the stream to be 7.2 Gyr, based on a drift rate at a velocity equal to the cluster dispersion of 1.1 \kms\ and a length of the analyzed region of 200 pixels, or, 8.1 kpc.  The average age of the stream is therefore 3.6 Gyr.   Consequently the average rate of gap creation is 0.17 gaps kpc$^{-1}$ Gyr$^{-1}$ for the Pal 5 trailing stream and we recommend a conservative error of a  factor of 2 on this rate. In an earlier paper \citep{Carlberg:12} we used the age of the oldest part of a set of streams to derive the gap rate. Modifying this to the average rate increases the gap rates a factor of two.  Revised gap rates for the available streams using the mean age are given in Table~1.

Sub-halos in galactic halos create gaps in a stellar streams at a rate which depends on the smallest visible gap which depends on the width, $w$, of the stream, as worked out in \citet{Carlberg:12}. The earlier calculations were oriented towards structures further out in the halo and were scaled to 100 kpc.  For this paper we undertook additional cold stream simulations that concentrated on the 15-30 kpc range and doubled the sub-halo orbit sampling. Since the gap width has a significant mass and radius dependence the simple fits to the results have a somewhat steeper slope with stream width,  
\begin{equation}
{\mathcal R}_\cup(w,r) = 5.0\times10^{-4} {n(r)\over{n_0}} r_{30}^{0.45}
w^{-1.16}\, {\rm kpc}^{-1} \,{\rm Gyr}^{-1}, 
\label{eq_w_rate}
\end{equation}
for $w$ in kpc, $r_{30}$ the orbital radius normalized to 30 kpc, and the constant evaluated for a stream age of 3.5~Gyr. For Pal~5's galactocentric distance of 19 kpc $n(r)=22 n_0$ from Figure~11 of \citet{Aquarius}. For widths, $w$ greater than a few kpc,  Equation~\ref{eq_w_rate} is similar to the result of \citet{Carlberg:12}, but the steeper slope predicts relatively more gaps in thin streams at small galactocentric radius than the same stream at a larger radius. 

The relation of Equation~\ref{eq_w_rate} is developed from a completely cold, zero-width stream. The width in the relation arises from the assumption that the smallest gap length that is visible is the width of stream, on the basis that the random motions of the stars in the stream are well described by epicycles which have an extent perpendicular to the steam nearly equal to the extent along the stream \citep{BT:08}.  The cold stream assumption runs into a problem for low mass halos where the width of the stream becomes significantly greater than the scale radius of the sub-halo, beyond which stream stars  are not sufficiently perturbed  to create a visible gap.  For the Pal~5 stream of 0.12~kpc width and setting $w=aR_s(M)$ we find $M_w=1.1\times 10^6 (1/a)^{1/0.43}\msun$.  The mass is close to the minimum mass halos that produce the gaps and will lead to fewer visible gaps in warm stream simulations. A second diminishing effect is that there will be a small range in angular momenta \citep{EB:11}, which translates into mean guiding center radius  in the star streams  such that differential rotation will be an additional blurring effect that increases with distance down the stream. Consequently the cold stream rates will be upper limits to more detailed simulation results.

The  star stream gap-rate vs width relation appropriate for galactocentric radii less than 30 kpc is shown in Figure~\ref{fig_gapwidth} along with the available data. The dashed line is a variance weighted straight line fit. 
The fitted line has a slope of $-0.59\pm0.35$, which is about 1.6 standard deviations below the predicted value.
Adopting mean ages rather than end-point ages for the streams doubles the estimated gap rates and helps bring the data into agreement with the prediction.  

\section{Discussion and Conclusions}

We have presented improved data which extends the known length of the northern (trailing) Pal~5 tidal stream to nearly 23\degr. 
The primary focus of this paper is to introduce an objective and statistically quantifiable approach to counting gaps, in the stream. 
At 90\% confidence the results vary some 40\% depending on the gap filter shape, but at 95\% confidence and above the counts agree within one. 
At 99\% confidence we find 5 gaps, where we have discounted one gap as being in a region in which the data are compromised. 
Our statistical confidence in the existence of density variations in the stream is $\ge$99.9999\% on the basis of its $\chi^2$. 
Not surprisingly, identifying the location and width of gaps can only be done at lower statistical level, where we have settled on a relatively conservative value of 99\% confidence.  
Deeper images would be invaluable in improving the statistics. Or alternatively, velocities for all of the current stars (g$\simeq$21 mag) would be invaluable in improving the statistical signal.

The density structures within the first kiloparsec of Pal~5 are likely dominated by the orbital dynamics of stars escaping the cluster. However, at larger distances the rising number of gaps is consistent with the expectations of dark matter sub-halo induced gaps. The small radius, cold stream, gap rate in terms of the minimum mass sub-halo $M_8= M/10^8 \msun$  is,
\begin{equation}
{\mathcal R}_\cup(M_8,r) = 3.6\times 10^{- 4}{n(r)\over{n_0}} r_{30}^{0.26}
M_8^{-0.36}\, {\rm kpc}^{-1} \,{\rm Gyr}^{-1},
\label{eq_M_rate}
\end{equation}
at 3.5~Gyr. The total number of sub-halos above the minimum mass is $260 M_8^{-0.9}$.  User the mid point and lower end of the stream averaged  Pal~5 gap creation rate at $r=0.63$ gives $M_8=0.15-1.0\times 10^{-3}$ and  implies a total population of some $1.3-7.2\times 10^5$ sub-halos (over the 433 kpc normalizing volume) or 900-5000 inside 30 kpc, far above the numbers of visibly populated dark matter sub-halos in our galaxy. 
The inferred numbers are very sensitive to the rate estimates, varying as the 3.24 power.
An important limitation of our statistical counting of sub-halos is the use of completely cold stream in the simulations which give the most possible gaps at any sub-halo mass. The sub-halo masses we derive, about $10^6\msun$ are at the point where we expect that the stream width will reduce the number of persistent gaps.

To gain an approximate understanding of the numbers we take all the halos to have the same mass so that  ${\mathcal R}_\cup \simeq n \pi b v$, where $n$ is their volume density,  $v$ is the typical encounter velocity, and $b$ the largest distance of the typical (low mass) sub-halo that induces a gap. The value of $b$ will be comparable to the stream width, $\ell$, so we set $b=\ell/2$.
For Pal~5 ${\mathcal R}_\cup \simeq 0.17$ kpc$^{-1}$ Gyr$^{-1}$, and $v\simeq 220 \kms$.  We find $n\simeq 0.0041$ kpc$^{-3}$. If these are the mean numbers inside 30 kpc, then the total number inside this volume is about 1030 correcting the numbers for higher density at smaller radius with the Einasto mass function.  Assuming a total mass inside 30 kpc of $3\times 10^{11} \msun$ and 7\% of the mass in sub-halos gives a mean sub-halo mass of $2\times10^6\msun$ That is, the huge numbers of halos required to create the gaps is a simple consequence of the fact that the interaction distance for the smallest mass sub-halos that create gaps is comparable to the quite narrow stream width. The \citet{Aquarius} simulations find that sub-halos at these numbers and masses have a circular velocity near 1 \kms, not surprisingly about equal to the velocity dispersion in the stream. It should be noted that for the mass function found in the simulations the heaviest sub-halos dominate the total mass.

The main result of this paper is to put the counting of the gaps in the Pal~5 stream on more solid ground. The resulting gaps counts are reasonably statistically secure and their numbers supports the conclusion that the character and number of gaps in stellar streams is similar to what a LCDM cosmological structure model predicts.  In comparison to a cold stream the number of gaps found is about a factor of three lower than predicted, but given the substantial uncertainties in both gap finding and in the simplified theory of gap creation used here, this mainly suggests that both the data and the theory needs to be made more precise.  

\citet{Carlberg:09} stated: ``{\it If the lowest velocity dispersion streams are older than about 3 Gyr, then LCDM subhalos would be ruled out}'', noting that: ``{\it A firm conclusion will require more extensive orbit modeling in well matched simulations.}" \citet{Carlberg:12} provided extensive but idealized circular cold stream modeling of sub-halo interactions with very low velocity dispersion streams, reaching the conclusion that streams are eroded via gaps created at a linear rate at time.  
 Figures~\ref{fig_recon} and \ref{fig_meandensity} show that the Pal~5 stream has a fair fraction, say 50\% or so, of its stream removed in gaps. This stream is being eroded away and is harder to find as a continuous structure at large distance from the cluster.   The exact fate of low velocity dispersion streams will be very dependent on their orbit around the galaxy. If they avoid dense regions such as the disk and especially the bulge, the stars pushed out of the gap remain near to the stream in a sort of snowplow pileup. However, if the stream orbits close to the bulge then small orbital differences can quickly lead to wide physical separations.

Recent discussion of problems with the number of the most massive sub-halos \citep{BKBK:12,SW:12} has highlighted the sensitivity of the predicted sub-halo numbers to the overall mass normalization of the galaxy. A smaller overall galactic mass would somewhat improve the agreement of theory and observation for both the low and high mass end of the spectrum, but it will take additional physical process to resolve the discrepancy, one possibility being accounting for large scale flows \citep{BD:12}.  The concentration of satellites and streams to a polar plane \citep{LB:76,PPK:12} is an additional long standing puzzle that may call for a dwarf galaxy formation scenario with additional astrophysics \citep{Hartwick:09,KFdB:10}. We conclude that our indirect sub-halo counting technique, which relies only on the gravitational field of the sub-halos to induce star stream gaps, finds that the gap statistics requires a very large number of sub-halos, comparable to the numbers that LCDM cosmology predicts. 

Ultimately the use of stream gaps to constrain the properties of density sub-structures within dark halos, that is, sub-halos, depends on being able to reliably count gaps, age date the tidal stream, and then relate the rate of gap creation to the population of sub-halos. In this paper we have made some progress on the statistics of gaps and are able to rely on 99\% confidence gap detections. An unfortunate situation is that the inferred numbers are steeply dependent, the 3.2 power, of the gap creation rate. However, better stream modeling is entirely possible and the observational statistics will improve with deeper images and as velocity data becomes available.

\acknowledgements
This research is supported by NSERC and CIfAR.  We thank the referee, Adreas K{\"u}pper for constructive comments.

Funding for SDSS-III has been provided by the Alfred P. Sloan Foundation, the Participating Institutions, the National Science Foundation, and the U.S. Department of Energy Office of Science. The SDSS-III web site is http://www.sdss3.org/.

SDSS-III is managed by the Astrophysical Research Consortium for the Participating Institutions of the SDSS-III Collaboration including the University of Arizona, the Brazilian Participation Group, Brookhaven National Laboratory, University of Cambridge, Carnegie Mellon University, University of Florida, the French Participation Group, the German Participation Group, Harvard University, the Instituto de Astrofisica de Canarias, the Michigan State/Notre Dame/JINA Participation Group, Johns Hopkins University, Lawrence Berkeley National Laboratory, Max Planck Institute for Astrophysics, Max Planck Institute for Extraterrestrial Physics, New Mexico State University, New York University, Ohio State University, Pennsylvania State University, University of Portsmouth, Princeton University, the Spanish Participation Group, University of Tokyo, University of Utah, Vanderbilt University, University of Virginia, University of Washington, and Yale University.

{\it Facilities:} \facility{Sloan}

\end{document}